\documentclass[12pt]{article}
\newlength{\abstractwidth}
\usepackage{epsf}
\usepackage{color}

\renewcommand{\thefootnote}{\fnsymbol{footnote}}
\renewcommand{\thanks}[1]{\footnote{#1}}
\newcommand{\starttext}{
\setcounter{footnote}{0}
\renewcommand{\thefootnote}{\arabic{footnote}}}

\newcommand{\bea}{\begin{eqnarray}}
\newcommand{\eea}{\end{eqnarray}}
\newcommand{\ee}{\end{equation}}
\newcommand{\be}{\begin{equation}}

\def\cN{{\cal N}}
\def\cO{{\cal O}}

\def\half{ {1\over 2}}
\def\p{\partial}

\def\a{\alpha}
\def\b{\beta}

\def\g{\gamma}

\def\g{\gamma}

\def\ch{{\rm ch}}
\def\sh{{\rm sh}}

\def\no{\nonumber}
\def\sm{\smallskip}

\long\def\symbolfootnote[#1]#2{\begingroup%
\def\thefootnote{\fnsymbol{footnote}}\footnote[#1]{#2}\endgroup}

\begin{document}
\starttext
\setcounter{footnote}{0}

\baselineskip=16pt

\begin{flushright}
UCLA/08/TEP/29\\
CPHT-RR079.1008\\
23 October 2008 
 \end{flushright}

\bigskip

\begin{center}

{\Large \bf Exact Half-BPS Flux Solutions in M-theory II: }

\medskip

{\Large \bf Global solutions  asymptotic to $AdS_{7}\times S^{4}$}\symbolfootnote[2]{\noindent This work was supported in
part by NSF grants PHY-04-56200 and PHY-07-57702.

\medskip

\noindent
{\sl E-mail addresses}: 
dhoker@physics.ucla.edu; johnaldonestes@gmail.com; gutperle@physics.ucla.edu;\\
dk320@physics.ucla.edu.}

\vskip .5in

{\large  Eric D'Hoker$^{a}$, John Estes$^{b}$, Michael Gutperle$^{a}$ and  Darya Krym$^{a}$}

\vskip .2in

{$\ ^{a}$ \sl Department of Physics and Astronomy }\\
{\sl University of California, Los Angeles, CA 90095, USA}

\vskip .2in

\vskip .2in
 
{$\ ^{b}$\sl  Centre de Physique Th«eorique, Ecole Polytechnique,}\\
{\sl FÐ91128 Palaiseau, France}

\end{center}

\vskip .2in

\begin{abstract}

\vskip 0.1in

General local half-BPS solutions in M-theory, which have $SO(2,2)\times SO(4)\times SO(4)$ symmetry and are asymptotic to $AdS_{7}\times S^{4}$, were constructed in exact
form by the authors in [arXiv:0806.0605].  In the present paper, suitable regularity
conditions are imposed on these local solutions, and corresponding globally well-defined 
solutions are explicitly constructed.  The physical properties of these solutions are analyzed, 
and interpreted in terms of the gravity duals to extended 1+1-dimensional  half-BPS 
defects in the 6-dimensional CFT with maximal supersymmetry.

\end{abstract}

\baselineskip=16pt
\setcounter{equation}{0}
\setcounter{footnote}{0}


 \newpage

\section{Introduction}
\setcounter{equation}{0}

One realization of the  AdS/CFT correspondence 
\cite{Maldacena:1997re,Gubser:1998bc,Witten:1998qj} in M-theory is the duality  
of  the $AdS_{7}\times S^{4}$ vacuum and the 6-dimensional CFT which is obtained 
by a decoupling limit of the M5 brane world-volume theory 
\cite{Aharony:1998rm,Leigh:1998kt,Minwalla:1998rp}. The nonabelian  world-volume 
theory of multiple M5-branes is presently unknown and the 6-dimensional CFT has 
been formulated in the light cone gauge \cite{Aharony:1997an}.
One interesting class of deformations in this theory is given by the insertion of 
local half-BPS chiral operators, where half-BPS means the operators preserve 
sixteen of the thirty-two supersymmetries.  The gravitational duals of these 
operators are the half-BPS solutions of Lin, Lunin, and Maldacena \cite{Lin:2004nb}.

\sm

In our recent paper \cite{D'Hoker:2008wc} (see also \cite{Boonstra:1998yu, deBoer:1999rh,Yamaguchi:2006te,Lunin:2007ab} for earlier work), new exact solutions of 
11-dimensional supergravity were constructed which preserve sixteen of the thirty-two supersymmetries.
In addition, the solutions preserve  a $SO(2,2)\times SO(4)\times SO(4)$ bosonic 
symmetry.  Correspondingly, the 11-dimensional metric is constructed as a warped 
product of $AdS_{3}\times S^{3}\times S^{3}$ over a 2-dimensional base space $\Sigma$.  These solutions can be interpreted as the gravity duals of extended supersymmetric 
defects in the CFT.
The solutions are local in the sense that for a bosonic background,  the vanishing 
of gravitino variation as well as the bosonic equations of motion and Bianchi 
identities are satisfied point wise, except at possible singularities.

\sm 

In general, the local solutions of \cite{D'Hoker:2008wc} contain a large variety of 
solutions many of which contain singularities.  An important problem is to pick out 
solutions which are asymptotic to either $AdS_4 \times S^7$ or $AdS_7 \times S^4$ 
and everywhere regular so that the supergravity approximation is valid.  In particular 
this requires one to examine the global structure of the solutions.  Similar analysis 
have been carried out in Type IIB supergravity in 
\cite{Lin:2004nb,D'Hoker:2007xz,D'Hoker:2007fq}. An amazing result in all cases is 
that the regularity conditions in addition to the general local solution admit a 
superposition principle for the half-BPS objects in the theory.  We expect such a 
principle to emerge from the analysis here.
There are two distinct classes of solutions which were found in \cite{D'Hoker:2008wc}:

\sm

Case I contains solutions asymptotic to $AdS_{4}\times S^{7}$. The superalgebra 
of symmetries is $OSp(4^*|2) \times OSp(4^*|2)$.  The super Lie algebra $OSp(4^*|2)$ 
is a particular real form of $OSp(4|2)$.  In the classification of \cite{D'Hoker:2008ix} 
this solution is case IV of table 11.
M-theory on $AdS_{4}\times S^{7}$ is dual to a 3-dimensional CFT which is 
obtained by a decoupling limit of the world-volume theory of M2 branes. The local 
BPS solution is dual to $1+1$-dimensional  conformal defects in the 3-dimensional 
CFT, analogous to the half-BPS defect solutions obtained in type IIB string theory 
\cite{D'Hoker:2007xy,D'Hoker:2007xz} (see also \cite{Gomis:2006cu} for earlier work).

\sm

Case II contains solutions asymptotic to $AdS_{7}\times S^{4}$. The bosonic 
isometries together with the supersymmetries  form a superalgebra which is given 
by  $OSp(4|2,{\bf R} ) \times OSp(4|2,{\bf R})$. The supergroup  $OSp(4|2,{\bf R}) $ 
is a different real form of $OSp(4|2)$ than the one appearing in case I. In the 
classification of \cite{D'Hoker:2008ix} this solution is case VII of table 12.  
M-theory on $AdS_{7}\times S^{4}$ is dual to a 6-dimensional CFT which is 
obtained by a decoupling limit of the world-volume theory of M5 branes. The 
local BPS solution is dual to $1+1$-dimensional  conformal defects in the 
6-dimensional CFT, analogous to the half-BPS Wilson loop solutions obtained 
in type IIB string theory \cite{D'Hoker:2007fq} (see also 
\cite{Yamaguchi:2006te,Lunin:2006xr} for earlier work).

\sm

The local solutions  presented in  \cite{D'Hoker:2008wc}  are very similar for the 
case I and II as the underlying integrable system is the same. However,  the 
analysis of the regularity and the global structure is quite different. In this paper 
we will focus on the solution of case II. The analysis of the regularity and global 
structure for case I will be analyzed in a separate   paper \cite{dgkenew}.

\sm

The structure of the paper is as follows. In section \ref{localsol} the  features of 
the local solution for case II which are important for the present paper will be 
reviewed. In section \ref{regularity}  the boundary conditions on the solution 
implied by regularity are analyzed. A general solution which satisfies suitable  
boundary condition is constructed and it is shown that the solution is regular 
everywhere. In section \ref{discussion} the global structure of the solutions  
as well as its interpretation in terms of the dual 6-dimensional conformal field 
theory is discussed. In appendix \ref{appendixa} detailed proof of the regularity 
of our solution is presented.

\section{Summary of local solution}
\setcounter{equation}{0}
\label{localsol}

In this section we review the local half-BPS solution of   \cite{D'Hoker:2008wc}. 
Derivations and more calculational details can be found in that paper. The 11-dimensional 
metric is a fibration of $AdS_{3}\times S^{3}\times S^{3}$ over a 2-dimensional 
Riemann surface $\Sigma$,
\bea
\label{metricansatza}
ds^2 = f_1^2 ds_{AdS_3}^2 + f_2^2 ds_{S_2^3}^2 + f_3^2 ds_{S_3^3}^2 +  ds_{\Sigma}^2
\eea
The four form field strength is given by
\bea
\label{fluxansatz}
F_{4}= 
g_{1a}\;  \omega_{AdS_{3}} \wedge e^{a}+
g_{2a}\;  \omega_{S_2^{3}} \wedge e^{a}+
g_{3a}\; \omega_{S_3^{3}} \wedge e^{a}
\eea
Here  $\omega_{AdS_{3}}$ and $\omega_{S^{3}_{2,3}}$ are the volume forms on 
$AdS_{3}$ and $S^{3}_{2,3}$ respectively. In addition  $e^{a}, a=1,2$  is the vielbein 
on $\Sigma$.  It is always possible to choose local complex coordinates $w,\bar w$ on 
the Riemann surface $\Sigma$ such that the 2-dimensional metric in (\ref{metricansatza}) 
is given by
\bea
\label{rhofactor}
ds_{\Sigma}^2= 4 \rho^{2 }\; |dw|^2
\eea
The metric factors $f_1, f_2, f_3, \rho$, as well as the flux fields 
$g_{1a}. g_{2a}$, and $g_{3a}$  only depend on  $\Sigma$.

\sm

The Ansatz respects $SO(2,2)\times SO(4)\times SO(4)$ symmetry which can be 
interpreted as the symmetries of a 1+1-dimensional conformal defect in the 6-dimensional 
M5 brane CFT. The condition that 16 supersymmetries are unbroken is equivalent,
for a purely bosonic background, to the statement that the gravitino supersymmetry 
variation $\delta_{\epsilon }\Psi_{M}$ vanishes for 16 linearly independent 
supersymmetry variation parameters. 

\sm

In \cite{D'Hoker:2008wc} the BPS  conditions were solved, and  it was shown that the 
half-BPS solution is completely determined by the choice of a 2-dimensional Riemann 
surface $\Sigma$, a real harmonic functions $h(w,\bar w)$ on $\Sigma$ and a complex 
function $G(w,\bar w)$ which is a solution of  the following  linear equation,
\bea
\label{gequation}
\p_w G = \half (G + \bar G) \p_w \ln h
\eea
In order to express the local half-BPS solution  in terms of $G$ and $h$ it is useful to define the following quantity
\bea
W^{2}= -|G|^{4}-(G-\bar G)^{2}
\eea
The metric factors in (\ref{metricansatza}) are then given by
\bea\label{metricfactors}
f_{1}^{6}&=& 4 h^{2}{(1-|G|^{2})\over W^{4}} \Big( |G-\bar G|+ 2 |G|^{2}\Big)^{3} \no\\
f_{2}^{6}&=& 4 h^{2}{(1-|G|^{2})\over W^{4}} \Big( |G-\bar G|- 2 |G|^{2}\Big)^{3} \no\\
f_{3}^{6}&=&{ h^{2} W^{2}\over 16 (1-|G|^{2})^{2}}
\eea
The metric factor in (\ref{rhofactor}) is given by
\bea
\rho^{6}&=&{(\partial_{w} h\partial_{\bar w }h)^{3} \over 16 h^{4}} \big(1-|G|^{2}\big) W^{2}
\eea
The fluxes $g_i$ are defined by conserved currents as follows
\bea
\label{bi}
(f_1)^3 g_{1w} = - \p_w b_1
&=& 2 (j_w^+ + j_w^-)
\no\\
(f_2)^3 g_{2w} = - \p_w b_2 &=& - 2 (j_w^+ - j_w^-)
\no\\
(f_3)^3 g_{3w} = - \p_w b_3 &=& {1 \over 8} \, j_w^3
\eea
where the conserved currents can be expressed in a compact way by defining
\be
J_{w}= {h\over G+\bar G} \bigg( \bar G (G - 3 \bar G + 4 G \bar G^2) \p_w G + G (G + \bar G) \p_w \bar G \bigg)
\ee
and are given by
\bea
j_w^+ &=&
2  i   \,
J_{w}
\bigg((G - \bar G)^2 - 4 G^3 \bar G\bigg)    W^{-4}
\no\\
j_w^- &=& 2  \, G J_{w}
\bigg(-2 G \bar G + 3 \bar G^2 - G^2 + 4 G^2 \bar G^2 \bigg)
 W^{-4}
 \no\\
j_w^3 &=&
3  \p_w h \, {W^2 \over G (1 - G \bar G)}
-2 J_{w} \, {(1 + G^2) \over G  (1 - G \bar G)^2}
\eea

It was shown in \cite{D'Hoker:2008wc} that the equations of motion of   as well as the Bianchi identities are satisfied for a harmonic $h$ and a $G$ which solves (\ref{gequation}).

\begin{figure}[tbph]
\begin{center}
\epsfxsize=4.5in
\epsfysize=2.3in
\epsffile{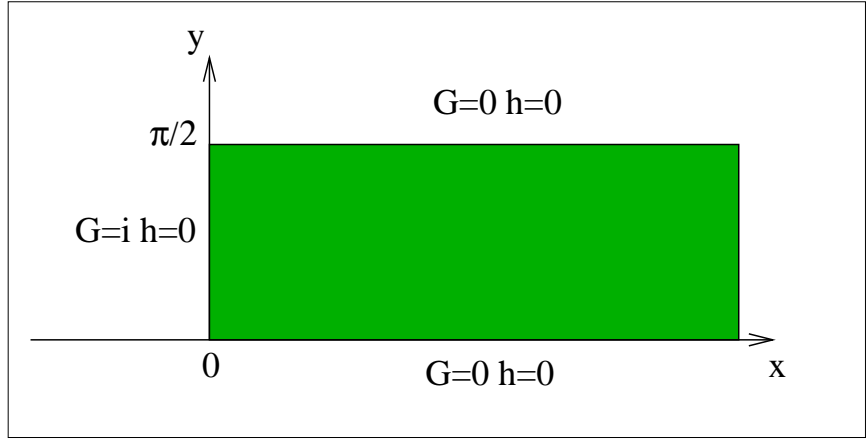}
\label{figure1}
\caption{$\Sigma$ and boundary conditions for $AdS_{7}\times S^{4}$ solution}
\end{center}
\end{figure}

The simplest solution is the maximally symmetric $AdS_{7}\times S^{4}$ itself.  The Riemann surface is the half strip $\Sigma=\{(x,y), x\geq0, 0\leq y\leq \pi/2\}$. Denoting the holomorphic coordinate as $w=x+i y$, the functions  $h$ and $G$ 
are given by
\bea\label{hgads}
h&=&-i\big(\cosh(2w)-\cosh(2\bar w)\big) \no\\
G&=&-i {\sinh(w-\bar w) \over \sinh(2 \bar w)}
\eea
Plugging this into  (\ref{metricfactors}) the metric factors become
\bea\label{adssolb}
f_1 = 2 \ch(x)
\qquad
f_2 = 2 \sh(x)
\qquad
f_3 = \sin(2y)
\qquad
\rho = 1
\eea
Note that the Riemann surface $\Sigma$ has three boundary components. The boundary is characterized by the vanishing of the harmonic function $h=0$.  Furthermore, taking $y = 0$ or $y = \pi / 2$, we find $G = 0$,
while taking $x = 0$, we find $G = + i$.  So $G$ takes the values $0$ and $+i$ on
the boundary of $\Sigma$.  The boundary of $AdS_7 \times S^4$ on the other
hand is located at $x = \infty$.

\section{Regularity}
\setcounter{equation}{0}
\label{regularity}
 
In this section, we analyze the regularity conditions and the global structure of the 
solution.  We look for solutions which are everywhere regular and which are 
asymptotic to $AdS_{7}\times S^{4}$. This leads to the following three assumptions 
for the geometry.

\begin{enumerate}
\itemsep -0.01in
\item  The boundary of the 11-dimensional geometry is asymptotic to
$AdS_7 \times S^4$.
\item   The metric factors
are finite everywhere, except at points where the geometry becomes
asymptotically $AdS_7 \times S^4$, in which case the $AdS_3$ metric factor
and one of the sphere metric factors diverge.
\item  The metric factors
are everywhere non-vanishing, except on the boundary of $\Sigma$, in which
case at least one sphere metric factors vanishes.  In addition, both sphere metric
factors may vanish only at isolated points.
\end{enumerate}
The second requirement guarantees that all singularities in the geometry are
of the same type as $AdS_7 \times S^4$.  The third requirement guarantees
that the boundary of $\Sigma$ corresponds to an interior line in the
11-dimensional geometry.

\sm

It follows from  (\ref{metricfactors}) that a particular combination of metric factors is 
very simple
\bea
\big(f_{1} f_{2} f_{3}\big)^{6}= h^{6}
\eea
The metric factor $f_{1}$ is positive definite and cannot vanish  \cite{D'Hoker:2008wc}.
Hence the  condition $h=0$ (which defines a 1-dimensional subspace in $\Sigma$)  
occurs if and only if at least one of the metric factors for the spheres $f_{2}$ or $f_{3}$ 
vanishes. It follows from  assumption $3$, that $h=0$ defines the boundary of $\Sigma$.

\sm

Note that the equation for $G$ (\ref{gequation}) is covariant under conformal reparamaterizations. This freedom allows one to choose local conformal 
coordinates\footnote{We use a slightly different notation: the coordinate $s$ in this 
paper was called  $x$ in \cite{D'Hoker:2008wc}.}
\bea
u=r+i \  s, \quad  \quad  r=h(w,\bar w) , \quad \quad s= \tilde h(w,\bar w)
\eea
Here, $\tilde h$ is the harmonic function dual to $h$ so that $u$ is holomorphic, 
i.e. $\partial_{\bar u}(h+i \tilde h)=0$. The domain of the new conformal coordinate 
$u$ is the right half plane and the boundary of $\Sigma$ is at $r=0$, i.e. the  vertical axis.

\sm

In the  coordinates $r,s$, it is useful to decompose $G$ into its real and imaginary parts,
\bea
G(r,s)  = G_r (r,s) + i G_s (r,s)
\eea
for $G_r,G_s$ real functions. The real and imaginary parts of   equation (\ref{gequation})
are respectively,
\bea
\label{bps4}
\p_r G_r + \p_s G_s & = & { G_r \over r}\label{geqone}
 \\
\p_r G_s - \p_s G_r & = & 0 \label{geqtwo}
\eea
Equation (\ref {geqtwo}) can be solved in terms of a single real potential
 \bea
\label{Psidef}
G_r  =  \p_r  ( r \Psi   ) \hskip 1in 
G_s   =  \p_s ( r \Psi   )
\eea
Equation (\ref {geqone}) becomes a second order partial
differential equation on $\Psi$,
\bea
\label{Psieq}
\Big ( \p_s^2 + \p_r ^2 + { 1 \over r} \p_r    - { 1 \over r^2} \Big ) \Psi (r,s)=0
\eea
The general local solution of (\ref{Psieq}) can be obtained by a Fourier transformation with respect to $s$, which produces an ordinary differential equation which can be solved by \cite{D'Hoker:2008wc}
\bea\label{fourierpsi}
\Psi(r,s)=\int_{-\infty} ^{\infty}{dk \over 2\pi}  \; \psi_2(k)K_1(kr) e^{-iks}
\eea
Here $K_1$ is  the modified Bessel function of the second kind. There is a second linearly independent solution of the form (\ref{fourierpsi}) which involves the modified Bessel function of the first kind $I_{1}(k r)$. However this solution has the wrong behavior for large $r$ and fails to obey the regularity condition $|G|^{2}\leq 1$.
 In \cite{D'Hoker:2008wc}  an explicit expression for $\Psi$ and $G$  was found by  defining  $C_2(v)$
\bea
C_2 (v) = \int _0 ^ \infty { dk \over 2 \pi} ~ \psi _2 (k) ~ e^{-kv}
\eea
and using the following integral representation of $K_{1}$
\bea\label{konedef}
K_1 (kr) = \int _1 ^\infty { t \, dt \over  \sqrt{t^2 -1 }}  \, e^{- tkr}
\eea
Using (\ref{fourierpsi})- (\ref{konedef}) $\Psi$ can be expressed in terms of $C_{2}$
\bea
\label{Psi3}
\Psi (r,s) = \int _1 ^\infty { t \, dt \over  \sqrt{t^2-1}}  \bigg ( C_2 (tr+is) + C_2 (tr+is)^* \bigg )
\eea
and (\ref{Psidef}) can be used to write $G$ as follows
\bea\label{gasintegral}
G  (r,s) =
r \int _1 ^\infty { dt \over \sqrt{t^2-1}} \bigg (
(1-t) C_2 ' (tr+is) + (1+t) C_2'(tr+is)^* \bigg )
\eea

\subsection{Boundary conditions on $G$}
\label{boundconsec}

In this section we analyze the boundary conditions $G$ has to satisfy at $h =0$ 
or in the $r,s$ coordinates at $r=0$ in order for the solution to be regular near 
the boundary. It will be useful to have the following expressions for the metric 
factors obtainable from (\ref{metricfactors})
\bea
\label{f2f3}
(f_1 f_2)^3 &=& - 4 r^2 {(1 - |G|^2) \over W}
\no\\
f_3^3 &=& - {r W \over 4 (1 - |G|^2)}
\eea
The behavior near the boundary (the regularity in the bulk of $\Sigma$ will be 
discussed in the next section) is exhibited by expanding $G_{r},G_{s}$, at fixed $s$,  
in a power series in $r$,
\bea
\label{powers}
G_{r}&=& \g_{1} \;r + \g_{3}\; r^{3}+ \g_{5}\;r^{5}+ \cO (r^{7})
\no\\
G_{s}&=& \g_{0} + \g_{2} \;r^{2}+ \g_{4}\;r^{4}+ \cO (r^{6})
\eea
where the $\g_{i}$ are all functions of $s$.
Note that the reality of the solution implies that $G$ is bounded $|G| \leq 1$ 
and hence no negative powers of $r$ can appear in the series expansion 
(\ref{powers}). Equation (\ref{geqone}) and (\ref{geqtwo}) impose the vanishing 
of even/odd powers of $r$ in $G_{r}/G_{s}$ respectively. Furthermore these 
equations impose differential equations in $s$ between the different $\g_{i}(s)$ 
but these relations will not be needed in the following.  The following power series 
expansions will be useful in the following analysis, 
\bea
\label{Was1}
W^2 & =& 
4 \g_0^2 (1 - \g_0^2) 
+ 8  \big( \g_0 \g_2 - 2 \g_0^3 \g_2 - \g_0^2 \g_1^2 \big)  r^2 + \cO(r^4) 
\no \\
1-|G|^{2} &=& 
(1- \g_{0}^{2})-\big( \g_{1}^{2}+2 \g_{0} \g_{2} \big) r^{2} + \cO(r^4)
\eea
When $\g_{0}\neq \pm1$, we have $|G| \neq 1$ as $r\to 0$. 
From (\ref{f2f3}), the product $f_{1} f_2$ goes to zero as $r \rightarrow 0$ 
and the geometry will be singular unless $W \sim r^{2}$.  It follows from 
(\ref{Was1}) that one has to choose $\g_{0}=0$ in order to avoid a singularity. 
If this is the case, $f_{2}$ will remain finite, while $f_{3}\to 0$ so that 
the volume of the sphere $S^{3}_3$ tends to zero. Comparing with the 
$AdS_{7}\times S^{4}$ solution, this behavior is associated with the $y=0,\pi/2$ 
boundary component of  (\ref{hgads}).

\sm

When $\g_{0}=\pm 1$,  we will have $f_{2}\to 0$ while $f_{3}$ will
remain finite, as long as the conditions $\g_{1}\neq \mp 2 \g_{2}$ and 
$\g_{1}^{2}\neq \mp \g_{2}$ are satisfied. If this is the case, the volume of the  
sphere $S^{3}_2$ will tend to zero. Comparing with the $AdS_{7}\times S^{4}$ solution, 
this behavior is associated with the $x=0$ boundary component of  (\ref{hgads}).

\sm

In summary we have the following boundary conditions on G  and the metric factors
\bea
\label{iiibc0}
G (0,s) = 0 ~~ &\qquad \Leftrightarrow \qquad & 
f_2 \neq 0 \qquad f_3 = 0, \qquad {\rm Vol}(S_{3}^{3})\to 0
\no\\
G (0,s) = \pm i &\qquad \Leftrightarrow \qquad & 
f_2 = 0 \qquad f_3 \neq 0  \qquad \; {\rm Vol}(S_{2}^{3})\to 0
\eea

\subsection{General regular solution}
\label{secthreetwo}

We first parameterize the boundary conditions for $G$ at $r=0$ which leads to 
solutions satisfying regularity conditions 1 to 3, listed at the beginning of 
section \ref{regularity}. As we shall see, a further condition needs to be imposed 
to guarantee the absence of singularities in the bulk, namely the values of $G$ in the second 
line of (\ref{iiibc0}) will be restricted to be either all positive or all negative.

\begin{figure}[tbph]
\begin{center}
\epsfxsize=5in
\epsfysize=2.3in
\epsffile{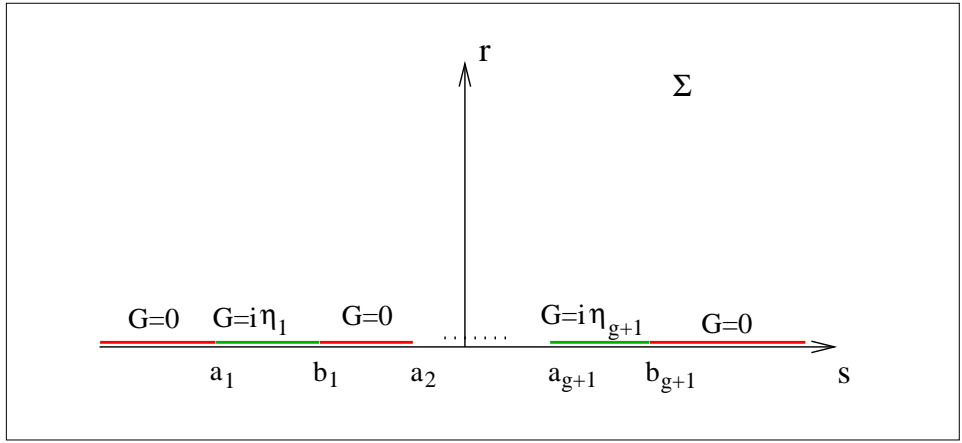}
\label{figure2}
\caption{The surface $\Sigma$ and boundary conditions for a general regular solution.}
\end{center}
\end{figure}

General boundary conditions for $G$ are given by the choice of $g+2 $ points on 
the $s$ axis,
\bea
\label{interv}
-\infty< a_1 < b_1 <    a_{2} < b_{2}< \cdots < a_{g+1} < b_{g+1}  <\infty
\eea
The $AdS_7 \times S^4$ solution corresponds to $g = 0$.  There are two kinds 
of intervals  which are distinguished by the boundary condition for $G(0,s)$.
 \bea
s\in [-\infty , a_{1}]& \qquad &G(s,0) =0\no\\
s \, \in \, [a_{n} \, , \, b_{n}] &  \;\;\qquad &G(s,0)= i \eta _n, \qquad n=1,2,\cdots g
\no \\
s \in [b_{n}, a_{n+1}] &  \qquad &G(s,0) =0, \;\;\;\qquad n=1,2,\cdots g\no\\
s\in [b_{g+1} , \infty ]&\qquad &G(s,0) =0
\eea
where $\eta_n = \pm 1$.  The boundary conditions can be implemented as follows:
\bea
\label{iiibc}
G(s,0) = \sum_{n=1}^{g+1} i \eta_n \bigg( \Theta(s - a_n) - \Theta(s - b_n) \bigg)
\eea
where $\Theta(s)$ is the step function. Since on each segment, $G(s,0)$ can take 
only the values $0, \pm i$, no two bumps can ``overlap", and this forces the $a_n$ 
and $b_n$ to alternate as in (\ref{interv}).

\sm

It remains to calculate $\psi _2 (k)$ from the boundary condition
in (\ref{iiibc}). To this end, we first take the Fourier transform
in $s$ with $ \ell >0$, of (\ref{Psidef}) with (\ref{fourierpsi}) plugged in
\bea
\int _{-\infty} ^ \infty ds ~ e^{ + i \ell s} ~ \p_s \Big ( r \Psi (s,r) \Big ) =
- i  \ell r \psi _2 (\ell) K_1 (\ell r)
\eea
whose $ r\to 0$ limit is simply obtained using the asymptotics of $K_1$,
and we find,
\bea
\int _{-\infty} ^ \infty ds ~ e^{ + i \ell s} ~ \lim _{r \to 0} \p_s \Big ( r \Psi (s,r) \Big ) =
- i   \psi _2 (\ell)
\eea
Using the  boundary condition of (\ref{iiibc}), we thus have
\bea
\label{psi2boundary}
\psi _2 (\ell) =\sum _n i \eta _n \int _{b_n} ^ {a_n} ds ~ e^{i \ell s}
=  \sum _n \eta _n { e^{i \ell a_n } - e^{i \ell b_n } \over \ell}
\eea
The Fourier transform can be done exactly
\bea
C_2(v) = - { 1 \over 2 \pi} \sum _n \eta _n \, \ln \left ( { v - i a_n \over v - i b_n} \right )
\eea
The expression for $G(s,r)$ may then be obtained using  the integral representation (\ref{gasintegral})
\bea
G(s,r) = \int _1 ^ \infty { dt \over \sqrt{ t^2 -1} } \bigg  (
r(1-t) C_2'(tr+is) + r (1+t) C_2'(tr +is )^* \bigg )
\eea
After some simplifications, we obtain,
\bea
G(s,r) = - { 1 \over 2\pi} \sum _{n=1}^{g+1} \eta _n \int _1 ^ \infty { dt \over \sqrt{ t^2 -1} }
\left [
{1 + i \alpha _n \over t + i \alpha _n} + {1 + i \alpha _n \over t - i \alpha _n}
- {1 + i \beta _n \over t + i \beta _n} - {1 + i \beta _n \over t - i \beta _n}  \right ]
\eea
where the quantities $\alpha_{n}$ and $\beta_{n}$ are defined as
\bea
\alpha _n &=&  (s - a_n)/r,\qquad \beta _n =  (s - b_n)/r,\quad n=1,2,\cdots g+1
\eea
 are both real.
Next we use the integral formula
\bea
\int_1^\infty {dt \over \sqrt{t^2 - 1}}
\bigg( {1 \over t + z} + {1 \over t - z} \bigg)  = { \pi \over \sqrt{1- z^2}}
\eea
The result is an algebraic expression for $G$ which solves (\ref{bps4}) and satisfies the boundary condition (\ref{iiibc}) and is given by
  \bea\label{gsolu}
G(s,r) &=&  -  \half \sum_{n = 1}^{g+1}  \eta_n
\left (
{ r+i s - i a_n \over \sqrt{ r^2 + (s-a_n)^2} }
- { r + i s - i b_n \over \sqrt{r^2 + (s-b_n)^2} } \right )
\eea
It is easy to see that (\ref{gsolu}) is equal to (\ref{iiibc}) in the limit $r\to 0$.

\subsection{Regularity in the bulk}

In the remainder of this section we give an argument that the general solution (\ref{gsolu}) is regular everywhere. Note that the general solution was constructed in section \ref{secthreetwo} by demanding that the geometry is regular at the boundary of $\Sigma$.

\sm

The general solution (\ref{gsolu}) should also approach the $AdS_{7}\times S^{4}$ boundary asymptotically as $r\to \infty$. The expression  for  $AdS_{7}\times S^{4}$. given in (\ref{hgads}) can be recovered from the general regular solution (\ref{gsolu})    by setting $g=0$,  $\eta_1 = -1$ , $a_1 = -2$ and $b_1 = 2$.
\bea\label{adssol}
G(s,r) &=&
{ r+i s +2  i  \over 2 |r+is + 2i| }
- { r + i s - 2i  \over 2 |r+is -2i| }
\no\\
&=& - i {\sh(w - \bar w) \over \sh(2 \bar w)}
\eea
where we have used the coordinates $r =  2 \sin(2 y) \sinh(2x) $ and
$s = 2 \cos(2y) \sinh(2x)$. The boundary of $AdS_{7}$ is reached by taking $x\to \infty$. Using the same coordinate change for the general solution it is easy to see that (\ref{gsolu})  approaches $AdS_{7}\times S^{4}$ in the limit $x\to \infty$.

It remains to show that the geometry is regular in the interior for $\Sigma$.
Since $h=r$ in the chosen coordinate system the regularity of the solution
requires that away from the boundary $W$ is required to satisfy the strict inequality $W^2 > 0$. Note that this
condition automatically guarantees that we also have $1-|G|^{2} > 0$.
Furthermore the relation
\bea
W^2 = - 4 |G|^4 - (G- \bar G)^2
= (|G-\bar G| - 2 |G|^2) (|G-\bar G| + 2 |G|^2)
\eea
shows that if $W^{2}>0$ then individually $(|G-\bar G| - 2 |G|^2) \neq 0$ and 
$(|G-\bar G| + 2 |G|^2) \neq 0$.  Examining the explicit formula for the metric 
factors (\ref{metricfactors}) one can see that they are then finite and the geometry 
is thus regular. It is shown in Appendix \ref{appendixa}  if the $a_{n},b_{n}$ 
obey the ordering (\ref{interv}) and  the $\eta_{n}$ are all either $+1$ or $-1$, 
that $W^{2}>0$ in the upper half plane and hence the solution is regular everywhere.

\sm

To see the converse, that is if $W^2 = 0$ in the bulk then the solutions is singular, 
we first note that for $W^2$ to vanish, either $(|G-\bar G| - 2 |G|^2)$ or 
$(|G-\bar G| + 2 |G|^2)$ must vanish.  Taking the ratio of $f_1$ and $f_2$ 
in (\ref{metricfactors}) we see that one of them must either be vanishing  
or be infinite resulting in a singular geometry.

\subsection{The $g= 1$ solution}
\label{genusonesec}

In this section, we examine the $g=1$ solution in detail, specifically we choose the parameters for the general solution (\ref{gsolu}) as follows:  
\be
\label{genusone}
g=1, \qquad a_1 = -2, \quad  b_1 = -1,\quad  a_2 = 0,\quad  b_2 = 1
\ee
In Figure 3, we show the behavior of the metric factors. The sphere metric factors 
alternatingly vanish as $r \rightarrow 0$.  
While as $r \rightarrow \infty$, the metric factors flatten out to those of $AdS_7 \times S^4$.  

\begin{figure}[tbph]
\begin{center}
\epsfxsize=5.2in
\epsfysize=3.8in
\epsffile{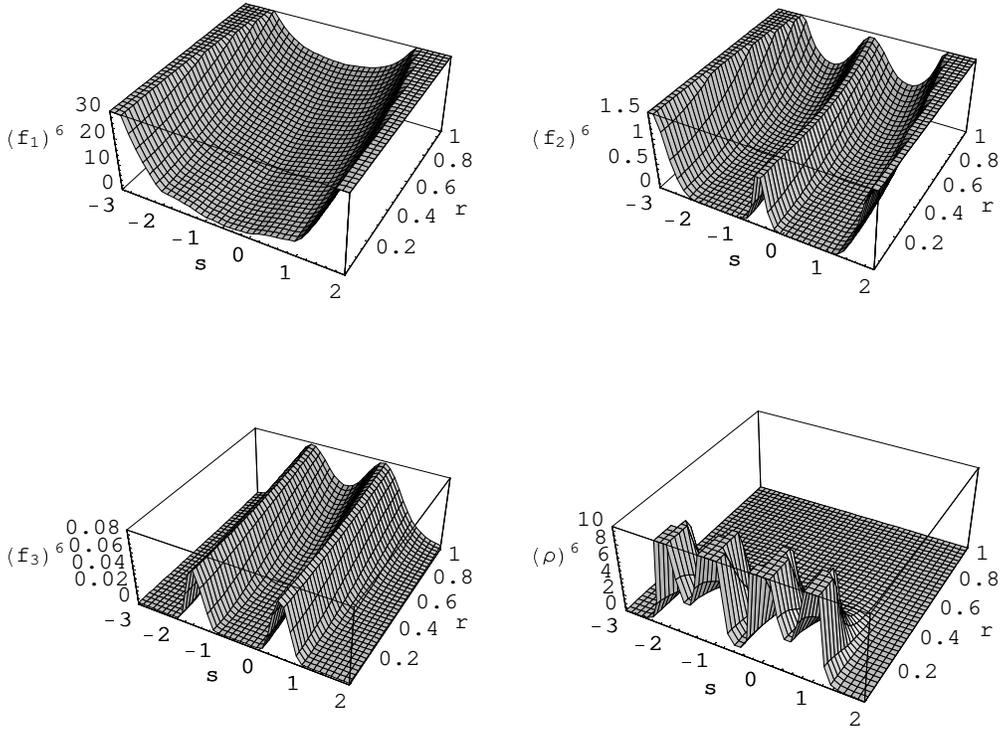}
\label{figure3}
\caption{Metric factors for a $g=1$ solution.}
\end{center}
\end{figure}
For $\rho$ there are singularities as we approach $r \rightarrow \infty$, but these are coordinate singularities and are due to the conformal transformation we made in order to map the half-strip to the upper half plane.

\begin{figure}[tbph]
\begin{center}
\epsfxsize=4.4in
\epsfysize=2.8in
\epsffile{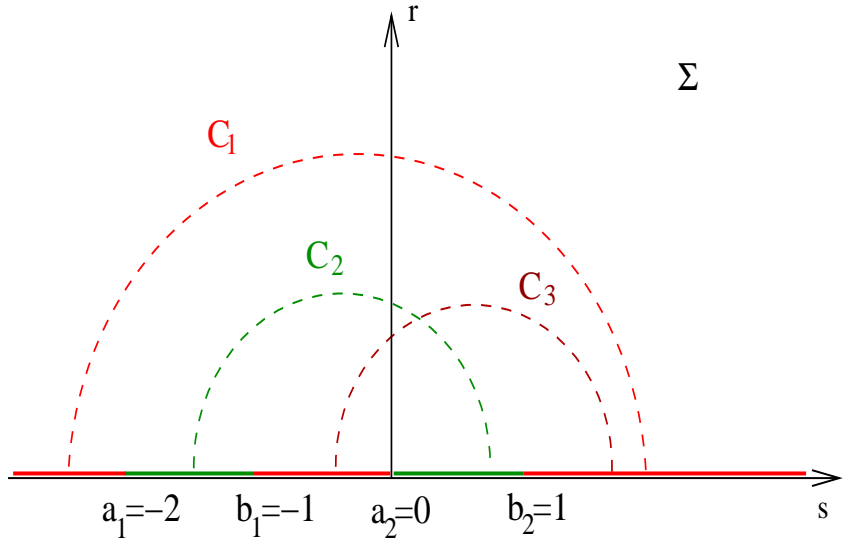}
\label{figure4}
\caption{Nontrivial four cycles for the $g= 1$ solution}
\end{center}
\end{figure}

An important feature of the solutions with $g>0$ is that there are additional nontrivial four cycles in the geometry. We can illustrate this feature for the $g=1$ solution (\ref{genusone}). In addition to the four cycle $C_{1}$ which is already present in the $g=0$ solution, there are two additional nontrivial four cycles $C_{2}$ and $C_{3}$.

The behavior of the fluxes for the $g= 1$ solution is very interesting. For comparison purposes we first plot the fluxes (\ref{bi}) for the  $AdS_7 \times S^4$ solution given by (\ref{gsolu}) with 
\be
g=0, \quad a_{1}=-2,\quad  b_{1}=1
\ee
\begin{figure}[tbph]
\begin{center}
\epsfxsize=3.0in
\epsfysize=2.0in
\epsffile{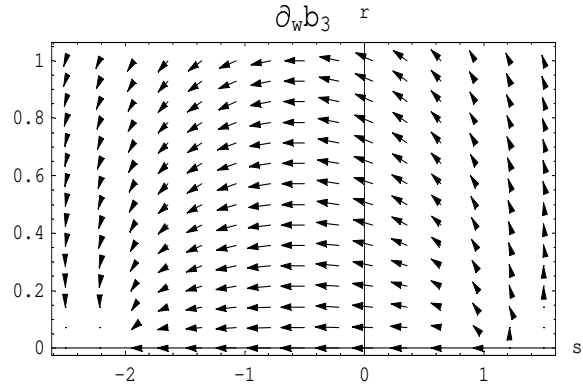}
\label{figure5}
\caption{The fluxes for a $g= 0$ solution.}
\end{center}
\end{figure}
Note that the fluxes $g_1$ and $g_2$ vanish identically  and the only nontrivial flux is $g_{3}$.  There is only one nontrivial topological cycle forming a four sphere. The integrated flux $g_{3 }$ is nothing but the non-vanishing four form flux through the four sphere in $AdS_{7}\times S^{4}$.

\begin{figure}[tbph]
\begin{center}
\epsfxsize=3in
\epsfysize=5.0in
\epsffile{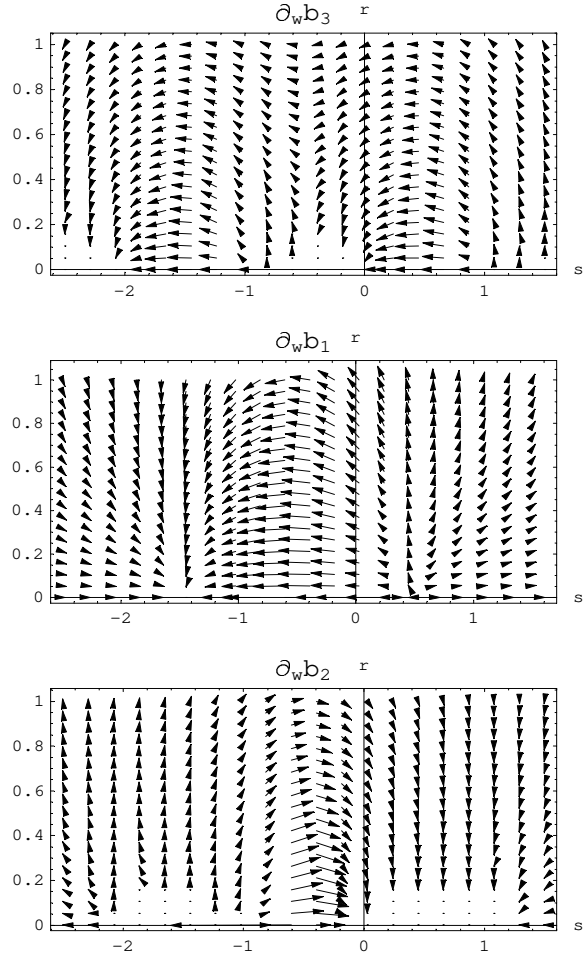}
\label{figure5}
\caption{The fluxes for a $g= 1$ solution.}
\end{center}
\end{figure}

In Figure 6, we plot the fluxes  for the $g=1$ solution (\ref{genusone}). Due to the complicated form of the currents  (\ref{bi}) we have not been able to integrate them to analytically obtain a closed form of the fluxes. However it is clear from figure 6 that the $g=1$ solution has indeed nontrivial flux through the cycles $C_{2}$ and $C_{3}$ for $g_{1}$ and $g_{2}$ respectively.

\section{Discussion}
\setcounter{equation}{0}
\label{discussion}

In the previous section we found a family of regular half-BPS solutions labelled by an 
integer $g$  and $2g+1$ real moduli. In this section we discuss the interpretation of 
these solutions from the point of view of the AdS/CFT correspondence. 
The  $AdS_{7}\times S^{4}$ spacetime is obtained as the near horizon limit of  a large 
number of M5  branes. The AdS/CFT duality relates M-theory on this background to 
the decoupling limit of the M5-brane world-volume theory which defines a 6-dimensional 
CFT with $(2,0)$ supersymmetry \cite{Leigh:1998kt,Aharony:1998rm}. 

\sm

A first step towards interpreting the solution is to understand the boundary structure. 
The only region on $\Sigma$ where  the spacetime becomes asymptotically 
$AdS_{7}\times S^{4}$ is  $r\to \infty$.   There is however another boundary 
component since the $AdS_{3}$ factor also has a boundary. This can be seen by 
rewriting the metric (\ref{metricansatza}).
\bea
\label{metricansatz}
ds^2 = {1\over z^{2}} \Big( f_1^2 ( dz^{2}+ dx^{2}-dt^{2})  + z^{2} f_2^2 ds_{S_2^3}^2 + z^{2 } f_3^2 ds_{S_3^3}^2 +  z^{2} ds_{\Sigma}^2\Big)
\eea
The boundary of $AdS_{3}$ is reached as $z\to 0$ and the boundary metric is obtained 
by stripping off the (divergent) conformal factor $1/z^{2}$. The $z^{2}$ factor in front of 
the metric factors of the spheres and $\Sigma$ implies  that the boundary in the limit 
$z\to 0$ is $1+1$-dimensional space extending in the $t,x$ plane.   The $SO(2,2)$ 
isometry of the $AdS_{3}$ factor corresponds to the conformal symmetry of the 
$1+1$-dimensional defect, which is contained in  the $OSp(4|2,{\bf R}) \times OSp(4|2,{\bf  R})$ supergroup of preserved  superconformal symmetries. Note that for all values of $g$ 
and the moduli there is only one defect.

\sm

The interpretation of the 1+1-dimensional half-BPS defect from the perspective of the 
dual  CFT is the supersymmetric self dual string solution of the 6-dimensional  $(2,0)$ supersymmetric M5-brane world-volume theory \cite{Howe:1997fb,Aganagic:1997zq,Bandos:1997ui} which was constructed in \cite{Howe:1997ue}.  The selfdual string in the $(2,0)$ 
theory  can also be interpreted as the boundary of an open M2 brane which ends on the 
M5-brane \cite{Strominger:1995ac,Dijkgraaf:1996cv} Unfortunately the action for 
multiple membranes is not well understood and the selfdual string soliton solution 
has only been derived for the abelian case of a single 5-brane.

\sm

There is a strong analogy of the selfdual string defect  with the BPS-Wilson loop in 
Type IIB string theory. While the details of the supergravity solution are somewhat 
different the general structure of the half-BPS flux solution and its moduli space 
presented in  section \ref{secthreetwo}  is intriguingly  similar to the Type IIB supergravity 
flux solutions dual to BPS Wilson loops which was found in  \cite{D'Hoker:2007fq}.

\sm

The BPS Wilson loop in $AdS_{5}\times S^{5}$ also has a probe description. The original proposal \cite{Maldacena:1998im,Rey:1998ik} identified the Wilson loop in the fundamental representation with a fundamental string with $AdS_{2}$ world-volume inside $AdS_{5}$. 
BPS-Wilson loops  in higher rank symmetric  representation and are identified with a 
probe D3 brane with electric flux with $AdS_{2}\times S^2$  world-volume inside $AdS_{5}$.     BPS-Wilson loops  in higher rank anti-symmetric  representation and are identified with 
a probe D5 brane with electric flux with $AdS_{2}\times S^4$  world-volume inside 
$AdS_{5}\times S^5$  \cite{Gomis:2006im,Gomis:2006sb}.   
 
\sm

The 1+1-dimensional BPS defect in the 6-dimensional CFT   can be viewed as the insertion  
``Wilson surface''-operator 
\cite{Ganor:1996nf,Berenstein:1998ij,Corrado:1999pi,Gustavsson:2004gj}. In the probe approximation one can use a analogy between the Wilson loop in $N=4$ SYM and the Wilson surface operators: The fundamental string is related to M2-brane probe. The D3 brane with electric flux and $AdS_{2}\times S^{2}$ world-volume is related to a probe M5 brane with 3-form flux on its $AdS_{3}\times S^{3}$ world-volume (with the $S^{3}$ embedded  in the $AdS_{7}$). The D5 brane with electric flux and $AdS_{2}\times S^{4}$ world-volume is related to  a probe M5 brane with three form flux on its $AdS_{3}\times S^{3}$ world-volume (with the $S^{3}$  embedded in the $S^{4}$).  These probe branes and their supersymmetry where analyzed in  \cite{Berman:2001fs,Chen:2007ir,Lunin:2007ab}

\sm
 
  The supergravity  solutions  we have obtained are  the analog of the ``bubbling" Wilson 
  loop solutions \cite{D'Hoker:2007fq,Yamaguchi:2006te,Gomis:2006sb}. They are fully backreacted and replace the probe branes by geometry and flux.  In particular as the discussion of the $g=1$ solution in section \ref{genusonesec} showed  there are  two new nontrivial four cycles $C_{2,3}$  
  in the $g=1$ solution. The fluxes through these cycles are  the remnants of the probe 
  M5-branes in the backreacted solution.
  
\sm

Unfortunately the $(2,0)$ theory for multiple M5-branes is not as  well understood as $\cN=4$ SYM theory. It is possible that the bubbling solutions can be useful in the understanding of the M5-brane theory. It would be interesting to see whether there is an analog of the matrix model description  of the BPS-Wilson loops (and its relation to the bubbling supergravity solution) for the Wilson surfaces. 
 
\sm

The general solution we have obtained has only one asymptotic $AdS_{7}\times S^{4}$ region. It would be interesting to  investigate whether its possible to have more than one asymptotic $AdS$ region, this would presumably correspond to a harmonic function $h$ with multiple poles. A similar phenomenon occurs in the case of half-BPS solutions which are asymptotic to $AdS_{4}\times S^{7}$ which we are currently investigating \cite{dgkenew}.

\bigskip

\noindent {\large \bf Acknowledgments}

\medskip

MG gratefully acknowledges the hospitality  of the International Center for 
Theoretical Science at the Tata Institute, Mumbai and  the Department of Physics 
and Astronomy, Johns Hopkins University during the course of this work.

 \newpage

\appendix

\section{Proof of the regularity condition $W^{2}>0$}
\label{appendixa}
\setcounter{equation}{0}

In this appendix  we shall prove a theorem which is central to establishing 
the regularity of the general solution constructed in section \ref{secthreetwo}.

\medskip

\noindent
{\bf Theorem 1} ~
{\sl When all $\eta_n$ are equal to one another,  the function $G$, defined by
\bea
G = -  \half \sum_{n = 1}^{g+1}  \eta_n
\left (
{ 1+i \a_n \over \sqrt{ 1 + \a_n^2} } - { 1 + i \b_n \over \sqrt{1 + \b_n^2} } \right )
\eea
satisfies $W^2 > 0$, for all $\a_n, \b_n$ subject to the ordering condition
\bea
\label{order1}
\a_1 < \b_1 < \a_2 < \b_2 < \cdots < \a_g < \b _g < \a_{g+1} < \b_{g+1}
\eea}
Numerical analysis suggests that this property holds, and also shows that, 
when not all $\eta_n$ are equal to one another, the condition $W^2 \geq 0$ is
violated for some range of $\a_n$ and $\b_n$. We shall prove Theorem 1 for 
$\eta _n = +1$ for all $n=1,\cdots, g+1$; the theorem for the opposite case 
$\eta _n=-1$ then follows immediately.

\sm

We begin by simplifying the condition $W^2 > 0$ as follows,
\bea
W^2 = - 4 |G|^4 - (G- \bar G)^2
= \left (|G-\bar G| - 2 |G|^2 \right ) \left ( |G-\bar G| + 2 |G|^2 \right )
\eea
The second factor on the right hand side of the last equality is manifestly positive for all $G$,
and may be dropped in the inequality. Thus, the condition $W^2>0$ becomes
equivalent to the condition $|G-\bar G| - 2 |G|^2 >0$, which is equivalent  to the 
following quadratic inequality
\bea
\label{Grim}
 X^2 + \left ( |Y| - \half \right )^2 < {1 \over 4} 
 \hskip 1in
 G= X + i Y 
\eea
where $X,Y$ are real.
In the sequel, it will be convenient to introduce the following notations,
\bea
p(\a) & \equiv & - {1 \over 2} \, {1 \over \sqrt{1+\a^2}}
\no \\
q(\a) & \equiv & + {1 \over 2} \, {\a \over \sqrt{1+\a^2}} \hskip 1in p^2+q^2={1 \over 4}
\eea
In terms of these functions, we define the following partial sums, for $m=1,2,\cdots, g+1$,
\bea
\label{XY}
X _m & = &
 \sum_{n = 1}^m  \Big ( p(\a_n)  - p(\b_n) \Big )
\no \\
Y_m & = &
 \sum_{n = 1}^m  \Big ( q(\b_n) - q(\a_n)  \Big )
\eea
so that the real and imaginary parts of $G$,  defined in (\ref{Grim}), 
are given by $X=X_{g+1}, Y= Y_{g+1}$.
 
\sm

The first key ingredient in the proof of Theorem 1 will be the fact that, for $\a \geq 0$,
the functions $p(\a)$ and $q(\a)$ are strictly monotonically increasing as $\a$ increases.
 
\subsection{The case $0 \leq \a_1$}

We begin by proving Theorem 1 for the following special ordering,
\bea
\label{order2}
0 \leq \a_1 < \b_1 < \a_2 < \b_2 < \cdots < \a_g < \b _g < \a_{g+1} < \b_{g+1}
\eea
Using the fact that  $p(\a)$ and $q(\a)$ are monotonically increasing with $\a$
for $\a \geq 0$,  it is immediate that $X_{g+1} < 0$ and $Y_{g+1} > 0$. 
Both bounds are sharp, as they can be saturated at the boundary of the domain 
(\ref{order1}) in the limit where $\a_n - \b_n \to 0$. A lower bound for
$X_{g+1}$ and an upper bound for $Y_{g+1}$ may be obtained by
letting $\beta _n - \a _{n+1} \to 0$ (with  $\a_{g+2} \equiv + \infty$).
Putting all together, we obtain the following double-sided bounds,
\bea
0 & < &  - X_{g+1}  < - p(\a_1)
\no \\
0 & < & Y_{g+1}  ~ ~ <  \half - q(\a_1)
\eea
To prove $W^2 >0$, we proceed 
recursively. Using the definition (\ref{XY}), we have,
\bea
X_{g+1} & = &
X_g + p(\a) - p(\b)
\no \\
Y_{g+1} & = &
Y_g + q(\b) - q(\a)
\eea
where we use the abbreviations $\a = \a_{g+1}$, and $\b= \b_{g+1}$.
Notice that we have $X_g < 0$ and $ Y_g >  0$.
The quantity of interest is
\bea
W_{g+1}^2 \equiv  X_{g+1}^2 + \left ( Y_{g+1} - \half \right )^2
\eea
Here, we have suppressed the absolute value sign on $Y_{g+1}$,
as we already know that $Y_{g+1} > 0$.
To show that $W^2 > 0$ holds, it will suffice to show that $W_{g+1}^2 < 1/4$.
Thus, we need to derive an optimal upper bound for $W_{g+1}^2$, and
show that this bound is less  than  $1/4$.

\sm

We first derive an upper bound on $W_{g+1}^2$ as a function of $\a$ and $\b$,
subject to the condition that $\b_g < \a < \b$. To this end, express $W_{g+1}^2$
as follows,
\bea
W_{g+1}^2 & = &
\left ( p(\b)  + x_g \right ) ^2
+ \left ( q(\b) - y_g \right ) ^2
\no \\
x_g & = & - X_g -p(\a) 
\no \\
y_g & = & - Y_g + \half + q(\a)
\eea
The bounds established earlier, namely $X_g < 0$ and $Y_g < 1/2$,
guarantee that $x_g > 0$ and $y_g > 0$ for all values of $\a \geq 0$.
We now search for the maximum of $W_{g+1}^2 $ as a function of $\beta$
over the interval $\beta \in [\alpha, +\infty]$, with $\alpha$ viewed as  fixed.
To determine it, we investigate the derivative with respect to $\beta$,
\bea
(W^2 _{g+1})'(\beta) = { x_g \beta - y_g \over \sqrt{1+ \b ^2}^3} 
\eea
This derivative can vanish in the interval $\beta \in [\alpha, +\infty]$
if and only if $x_g \alpha - y_g \leq 0$. If this is the case, the corresponding
point is $\beta _0 = y_g/x_g$, which should satisfy $\beta _0 > \a$.

\sm

Hence, the extrema of $W_{g+1}^2$ as a function of $\beta$ may be attained
either at $\beta = \beta _0$, or at either one of the extremities of the interval 
$\beta \in [\alpha, + \infty]$. These three values are given by, 
\bea
W_{g+1}^2 (\b_0) & = & x_g^2 + y_g ^2 + { 1 \over 4} - \sqrt{x_g^2 + y_g^2}
\no \\
W_{g+1}^2 (\a) & = &  x_g^2 + y_g ^2 + { 1 \over 4} + 2p(\a) x_g - 2q(\a)  y_g 
\no \\
W_{g+1}^2 (\infty) & = & x_g^2 + y_g ^2 + { 1 \over 4} - y_g
\eea
Since $x_g, y_g > 0$, it is manifest that $W_{g+1}^2 (\b_0) < W_{g+1}^2 (\infty)$.
Thus, $W_{g+1}^2 (\b_0)$ cannot be the optimal upper bound for $W_{g+1}^2(\beta)$.
Comparing the remaining two possible values, we find,
\bea
W_{g+1}^2 (\infty) - W_{g+1}^2 (\a)
= 2 p(\a)  X_g  + 2 \left ( 1 - q(\a) \right ) Y_g
\eea
Given that $X_g < 0$ and $Y_g > 0$, it follows that the right hand
side is positive and so that $W_{g+1}^2 (\infty)$ is the optimal upper bound.
In summary,
\bea
\label{ineq1}
W_{g+1}^2  & < & V_g (\alpha_{g+1})
\no \\
V_g(\a _{g+1}) & \equiv & \Big ( X_g + p(\a_{g+1}) \Big )^2
+ \Big (  Y_g - q(\a_{g+1}) \Big )^2
\eea
for all values of $\a_{g+1}$ such that $\beta _g < \a_{g+1}$.

\sm

Since $X_g < 0$ and $Y_g > 0$, it is straightforward to derive an upper
bound for the right hand side of (\ref{ineq1}). Indeed, both terms increase
as $\a_{g+1}$ decreases. Thus, the optimal bound for the right hand side is
attained when $\a_{g+1}$ assumes its smallest possible value, which is $\a_{g+1} = \b _g$.
Hence, we have
\bea
V_g (\alpha_{g+1} ) < V_g (\beta _g) \hskip 1in \alpha_{g+1} \in [\beta _g, \infty]
\eea
But, using the definitions of $X_g$ and $Y_g$ in terms of $\a_n$ and $\b_n$,
we see that the quantity $V_g (\beta _g)$ admits a drastic simplification,
\bea
V_g (\beta _g) = \Big ( X_{g-1} + p(\a_g) \Big )^2
+ \Big (  Y_{g-1} - q(\a_g) \Big )^2 = V_{g-1} (\a_g)
\eea
Combining all, we get a recursive series of bounds,
\bea
W_{g+1}^2 <  V_g (\alpha_{g+1})
< V_{g-1} (\a_g) < V_{g-2} (\a_{g-1}) < \cdots < V_0 (\a_1)
\eea
From their definitions, $X_0=Y_0=0$, we readily find $V_0(\a_1) = 1/4$, so that
$W_{g+1}^2 < 1/4$.
This concludes the demonstration of Theorem 1 for the case $0 \leq \a_1$.

\subsection{The case $\b_{g+1}\leq 0$}

Next, we proceed to proving Theorem 1 for the following special ordering,
\bea
\label{order3}
 \a_1 < \b_1 < \a_2 < \b_2 < \cdots < \a_g < \b _g < \a_{g+1} < \b_{g+1} \leq 0
\eea
It is not necessary to repeat the steps analogous to the proof for the
case $0 \leq \a_1$, since we can reduce the present case to 
the $\a_1 \geq 0$ case  by changing variables,
\bea
\a_n & = & - \tilde \b _{g+2-n} \hskip 1in n = 1, \cdots, g+1
\no \\
\b_n & = & - \tilde \a _{g+2-n}
\eea
The $\tilde \a_n$ and $\tilde \b_n$ are now all positive and satisfy the
following ordering,
\bea
0 \leq \tilde  \a_1 < \tilde   \b_1 < \tilde  \a_2 < \tilde  \b_2 < \cdots <
\tilde  \a_g < \tilde  \b _g < \tilde  \a_{g+1} < \tilde   \b_{g+1}
\eea
Denoting the corresponding function by $G^-$, and its real and imaginary parts
by $X_{g+1}^-$ and $Y_{g+1}^-$, we have by definition,
\bea
\label{XY1}
X _{g+1}^-
& = &
 \sum_{n = 1}^{g+1}  \left ( p(\tilde \b_n) - p(\tilde \a_n) \right )
= - \tilde X_{g+1}
\no \\
Y_{g+1}^-
& = &
 \sum_{n = 1}^{g+1}  \left ( q(\tilde \b_n) - q( \tilde \a_n)   \right )
= + \tilde Y_{g+1}
\eea
where $\tilde X_{g+1}$ and $\tilde Y_{g+1}$ are given by (\ref{XY}) but
with $\a_n \to \tilde \a_n$ and $\b_n \to \tilde \b _n$.
From the proof of the case $\a_1 \geq 0$, it now follows that $W^2 > 0$ 
also for this special case.

\subsection{The general case}

Next, we shall prove Theorem 1 for the cases whose ordering is given by
\bea
\label{order4}
 \a_1 < \b_1 <   \cdots < \a_N < \b _N \leq
 0 \leq \a_{N+1} < \b _{N+1} < \cdots < \a_{g+1} < \b_{g+1}
\eea
for $N=1, \cdots, g$. (The proof for the case with the ordering 
$  \cdots < \a_N < 0 < \b_N < \cdots $ follows the same steps, or may be 
derived by taking the limit $\a_{N+1}, \b_N \to 0$ in the ordering of
(\ref{order4}), and need not be detailed here.) 

\sm

It remains only to prove Theorem 1 for the ordering (\ref{order4}).
To do so, we use the fact that the behaviors of the variables
larger than 0 and those smaller than zero are independent of one another.
Concretely, we define the following partial sums,
\bea
\label{XY2}
X ^- & = &
 \sum_{n = 1}^N  \left ( p( \a_n) - p( \b_n) \right )
\no \\
X ^+ & = &
 \sum_{n = N+1}^{g+1}  \left ( p( \a_n) - p( \b_n) \right )
\no \\
Y^- & = &
 \sum_{n = 1}^N  \left ( q( \b_n) - q( \a_n )  \right )
\no \\
Y^+ & = &
\sum_{n = N+1}^{g+1}  \left ( q( \b_n) - q( \a_n) \right )
\eea
so that the full sums are given by
\bea
G = X_{g+1} + i Y_{g+1} & \hskip 1in & X_{g+1} = X^+ + X^-
\no \\
&& \, Y_{g+1} = Y^+ + Y^-
\eea
To the sums $X^+, Y^+$, we apply the results derived for case $0 \leq \a_1$,
while to the sums $X^-, Y^-$, we apply the results derived for case $\beta_{g+1}<0$,
namely
\bea
X^+ < 0 \qquad 0 < Y^+ < \half & \hskip 0.7in &
	(X^+)^2 + \left ( Y^+ - \half \right )^2 < { 1 \over 4}
\no \\
X^- >0 \qquad 0 < Y^- < \half &  &
	(X^-)^2 + \left ( Y^- - \half \right )^2 < { 1 \over 4}
\eea
From the fact that $X^+$ and $X^-$ have opposite sign, and the fact that
$Y^+ - 1/2$ and $Y^-$ have opposite sign, it follows immediately that
\bea
\left ( X^+ + X^- \right )^2 + \left ( Y^+ + Y^- - \half \right )^2 <
\left ( X^\pm  \right )^2 + \left ( Y^\pm   - \half \right )^2 \, < \, { 1 \over 4}
\eea
so that $|G-\bar G| - 2 |G|^2 > 0$, and thus $W^2 > 0$,
which completes the proof of Theorem 1 in the general case.
Note that the range of $G$ in the general case is all of the disc
$|G - 1/2|< 1/2$.

\end{document}